\documentclass[prl,showpacs,twocolumn]{revtex4}
\usepackage{amsfonts,graphics}

\begin{document}

\title{Using the Chern-Simons Action for a Self-Consistent Determination of the
Magnetic Screening Mass in Thermal QCD}

\author{G.M. \surname{von Hippel}}
\author{R.R. \surname{Horgan}}
\affiliation{Department of Applied Mathematics and Theoretical Physics,
University of Cambridge, Centre for Mathematical Sciences, Cambridge CB3 0WA, United Kingdom}

\preprint{DAMTP-2002-89}

\begin{abstract}
We present a self-consistent determination of the screening mass for
chromomagnetic fields in QCD within the framework of dimensional reduction.
The three-dimensional Chern-Simons density is used as a mass term for a
self-consistent perturbative calculation that yields a value of
$m\approx 1.604\frac{g^2N}{2\pi}T$ for the magnetic screening mass.
\end{abstract}

\maketitle

The perturbative expansion of thermal gauge theories is well known to be plagued
by infrared divergences that render perturbative computations of most quantities
impossible, as first noted by Linde \cite{linde:problem}. In order to overcome
these difficulties, several effective field 
theories have been proposed \cite{braaten:eft, kajantie:generic}. These are
based on the concept of dimensional reduction, where the static Matsubara modes
of the thermal theory are described by an effective theory in three-dimensional
Euclidean space, with the other modes (and any fermions) integrated out. For the
chromomagnetostatic modes of QCD, the lowest-order effective theory is the conventional
Yang-Mills theory in three dimensions, with dimensionful coupling
$g_3=g\sqrt{T}$.

In three dimensions, the dynamical generation of a gluon mass is expected.
Unfortunately, a perturbative evaluation of this mass is not possible, for
similar reasons as in the finite-$T$ case: The superrenormalisability of the
theory in three dimensions leads to infrared divergences beyond the one-loop
level, and no mass for the chromomagnetic modes is generated at one loop. In order to
determine the magnetic screening mass within the framework of perturbation
theory, one will therefore have to resort to self-consistent approximations.

In this paper, we show how the Chern-Simons Lagrangian can be used in the
context of a self-consistent resummation system, and discuss some of the
potential problems with our approach, and why some of these problem might
actually be considered as advantages.

\section{Methods}

The Chern-Simons Lagrangian in three (Euclidean) dimensions is given by
\cite{deser:tmym}
\begin{equation}
\mathcal{L}_{CS}=-im\epsilon^{\mu\nu\rho}\mathrm{tr}\left(A_{\mu}\partial_{\nu}A_{\rho}
+\frac{2g_3}{3}A_{\mu}A_{\nu}A_{\rho}\right)
\end{equation}
and is gauge-invariant (up to a total divergence) under small gauge
transformations. For invariance of the action under large gauge transformations
one needs the condition 
\begin{equation}
\frac{4\pi m}{g_3^2}\in\mathbb{Z}
\label{quant}
\end{equation}
For the purposes of doing perturbation theory, however, small gauge invariance
is all that is needed, as perturbation theory does not experience contributions
from non-trivial vacua. We will therefore regard $m$ as an arbitrary parameter
with no \emph{a priori} relation to $g_3$.

If a Chern-Simons term is added to the standard Yang-Mills Lagrangian, the gauge
fields acquire a mass from the quadratic part of $\mathcal{L}_{CS}$, and the
bare gluon propagator in a covariant gauge becomes
\begin{equation}
D^{ab}_{\mu\nu}(p)=\left(\frac{p^2\delta_{\mu\nu}-p_{\mu}p_{\nu}
-m\epsilon_{\mu\nu\rho}p^{\rho}}{p^2(p^2+m^2)}
+\xi\frac{p_{\mu}p_{\nu}}{(p^2)^2}\right)\delta^{ab}
\label{prop}
\end{equation}
while the three-gluon vertex receives a momentum-independent extra contribution
$igmf^{abc}\epsilon_{\mu\nu\rho}$ from the cubic part of the Chern-Simons
Lagrangian. The propagator is manifestly infrared-safe only in the Landau gauge
$\xi=0$, while in other covariant gauges, spurious infrared-divergences appear.
While a cautious handling should be able to remove these infrared divergences
from all physical quantities, we will not concern ourselves with those
subtleties, and work exclusively in Landau gauge for reasons of convenience.

It is easily apparent that the Chern-Simons Lagrangian is odd under parity,
hence adding it to a Yang-Mills action violates the parity invariance of the
latter. This implies that the mass can have either a positive or a negative
sign; as we are only interested in the position of the pole, which depends only
on the square of the mass, we are free to chose the mass to be positive in all
our calculations.

A self-consistent resummation scheme is normally based on adding a quadratic
(parity-even) mass term to the Lagrangian and resubtracting it as a counterterm
fixing the pole of the propagator at $p^2=-m^2$:
\begin{equation}
\mathcal{L}=\mathcal{L}_0+m^2\mathcal{L}_m-\delta m^2\mathcal{L}_m
\end{equation}
while demanding $\delta m^2=m^2$ in order not to change the original theory,
leading to a gap equation 
\begin{equation}
m^2=\Pi(-m^2)
\end{equation}
where $\Pi(p^2)$ is the perturbative shift of the inverse propagator from its
free form.

For the Chern-Simons mass term, which is only linear in the mass, a
slightly modified scheme is needed. The Lagrangian with the mass term added and
subtracted reads
\begin{equation}
\mathcal{L}=\mathcal{L}_0+m\mathcal{L}_{CS}-\delta m\mathcal{L}_{CS}
\end{equation}
where $\mathcal{L}_0$ is the standard Yang-Mills Lagrangian, and we demand
\begin{equation}
\delta m=m
\end{equation}
while considering $\delta m$ as a counterterm that fixes the pole of the
propagator at $p^2=-m^2$.

By adding the Chern-Simons term to the Lagrangian, we have changed the
admissible tensorial structure of the self-energy, which may now have a
parity-odd part:
\begin{equation}
\Pi_{\mu\nu}(p)=\left(p^2\delta_{\mu\nu}-p_{\mu}p_{\nu}\right)A(p^2)
+m\epsilon_{\mu\nu\rho}p^{\rho}B(p^2)
\end{equation}
where the parity-even part $A(p^2)$ can be isolated by contracting with a
symmetric tensor such as $\delta_{\mu\nu}-\lambda\frac{p_{\mu}p_{\nu}}{p^2}$,
while the parity-odd part $B(p^2)$ may be isolated by contraction with
$\epsilon_{\mu\nu\lambda}$.

Taking into account that $\delta m=m$, the inverse propagator turns out to be
\begin{equation}
\mathcal{D}^{-1}_{\mu\nu}(p)=\left(p^2\delta_{\mu\nu}-p_{\mu}p_{\nu}\right)(1+A(p^2))
+m\epsilon_{\mu\nu\rho}p^{\rho}B(p^2)
\end{equation}
such that the resummed propagator becomes
\begin{equation}
\mathcal{D}_{\mu\nu}(p)=\frac{\left(1+A(p^2)\right)\left(p^2\delta_{\mu\nu}-p_{\mu}p_{\nu}\right)
-B(p^2)m\epsilon_{\mu\nu\rho}p^{\rho}}
{\left[1+A(p^2)\right]^2p^2\left[p^2+\left(\frac{B(p^2)}{1+A(p^2)}m\right)^2\right]}
\end{equation}
To keep the pole of the propagator fixed at $p^2=-m^2$, we have to demand
\begin{equation}
\label{gap}
B(-m^2)=1+A(-m^2)
\end{equation}
as the appropriate gap equation for the self-consistent determination of the
magnetic mass with a Chern-Simons mass term.

\section{Results}

Adding the Chern-Simons Lagrangian does not change the number or topology of the
Feynman diagrams for the theory. The one-loop self-energy is therefore given by
the sum of the same three diagrams as in the non-massive pure gauge theory.

Let us first compute $B(-m^2)$. The only contributions to the parity-odd part of
the gluon self-energy come from the diagram with two three-gluon vertices,
because the external index structure of the other diagrams has not been changed
by the addition of the Chern-Simons term. The contraction of this diagram with
the $\epsilon$-tensor yields
\begin{widetext}
\begin{equation}
B(p^2)=\frac{g_3^2N}{p^2}\int\frac{d^3k}{(2\pi)^3}\frac{(k^2p^2-(k\cdot p)^2)
(5k^2+5k\cdot p+4p^2+2m^2)}{k^2(k^2+m^2)(k+p)^2((k+p)^2+m^2)}
\end{equation}
\end{widetext}
This integral is convergent both in the infrared and the ultraviolet.
For our purposes, we need to continue it to $p^2=-m^2$.
Since there is a threshold and corresponding cut in the complex plane
at $p^2=0$, coming from the $p^2$ term in the denominator of the
propagator (\ref{prop}), we have to assure the
analyticity of the integral by making sure that no poles of the
integrand cross the axis of integration. Since the problematic poles
are those from the $(k+p)^2$ term in the denominator of the integrand,
this can be done by rotating the axis of integration by the same phase
as $p$: For $p=e^{i\chi}m$, take $\arg(k)=\chi-\epsilon$, with
$\epsilon$ infinitesimally small. Using this prescription, we can
 evaluate the integral numerically as
\[
B(-m^2)=\Big(0.1677(3)+0.0160(5) i\Big)\frac{g_3^2N}{m}
\]
For an
analytic calculation this prescription is slightly awkward.
 We can, however,
still calculate the integral without recourse to any regulator by noticing that,
since $k$ and $\theta$ only enter the integral via the combinations $k^2$,
$\cos^2\theta$ and $k\cos\theta$, we can replace the
$\int_0^{\infty}dk\int_{-1}^1d\cos\theta$ integration by a
$\int_{-\infty}^{\infty}dk\int_0^1d\cos\theta$ one, which can be
performed analytically using the residue theorem. Alternatively, we
could use partial fractions for a 
straightforward evaluation of the integral (where, however, a regulator such as
dimensional regularisation needs to be employed).
Continuing $p\to im$, we finally find the value
\[
B(-m^2)=\left(\frac{1}{8}+\frac{27}{32}\log 3\right)\frac{g_3^2N}{2\pi m}
+i\frac{g_3^2N}{64m} 
\]
for the odd part of the self-energy at the pole, which agrees perfectly with the
numerical result obtained using the rotated contour.

The computation of $A(-m^2)$ is a bit more subtle. Superficially, $A(-m^2)$
contains linear divergences, and while these would disappear using e.g.
dimensional regularisation, it seems that the value of $A(-m^2)$ would be
dependent on the regularisation scheme used. Since $A(-m^2)$ enters the gap
equation (\ref{gap}), this would render the self-consistent mass
regulation-dependent, thus invalidating the entire procedure. Fortunately, the
situation is yet a bit more subtle: On dimensional grounds, it is known that the
factor multiplying $p_{\mu}p_{\nu}$ in the self-energy must be ultraviolet
convergent in three dimensions. Gauge invariance, however, implies that the
parity-even part of the self-energy is of the form
$\left(\delta_{\mu\nu}p^2-p_{\mu}p_{\nu}\right)A(p^2)$, and hence the
divergences in $A(p^2)$ must disappear for any gauge-invariant regulator, so as 
to render the value of $A(p^2)$ regulator-independent. We choose to evaluate
$A(-m^2)$ in a manifestly regulation-independent fashion by identifying the
self-energy integral multiplying $p_{\mu}p_{\nu}$, which can be isolated by
contracting with the projector $\delta_{\mu\nu}-3\frac{p_\mu p_\nu}{p^2}$.

The integral which then needs to be evaluated is
\begin{widetext}
\[
A(p^2)=\frac{g_3^2N}{4 p^4}\int\frac{d^3k}{(2\pi)^3}\frac{1}{k^2(k^2+m^2)(k+p)^2((k+p)^2+m^2)}\]\[
\times \left(  -8k^6p^2 + 24 k^4(k\cdot p)^2 +  8k^4(k\cdot p)p^2 + 36 k^2(k\cdot p)^3
-8k^4p^2m^2 - 14k^4p^4 + 24k^2(k\cdot p)^2m^2 +  66k^2(k\cdot p)^2p^2 \right.\]\[\left.
-12(k\cdot p)^4+14 k^2(k\cdot p)p^2m^2 + 12k^2(k\cdot p)p^4 + 18(k\cdot p)^3
-4k^2p^2m^4 - 4k^2p^6 + 12(k\cdot p)^2m^4 + 24(k\cdot p)^2p^2m^2 \right.\]\[\left.
+4(k\cdot p)^2p^4 +8(k\cdot p)p^2m^4 + 8(k\cdot p)p^4m^2\right)
\]
\end{widetext}
As above, we have to either rotate the
integration contour for a numerical evaluation yielding
\[
A(-m^2)=\Big(-0.0881(2)+0.0165(5) i\Big)\frac{g_3^2N}{m}
\]
or
perform the integral analytically (making use of a computer algebra program for
obvious reasons) and take the limit $p\to im$ afterwards to obtain
\[
A(-m^2)=\left(\frac{3}{8}-\frac{27}{32}\log 3\right)\frac{g_3^2N}{2\pi m}
+i\frac{g_3^2N}{64 m}
\]
which again agrees with the numerical result.

The self-energies of Chern-Simons massive three-dimensional Yang-Mills theory
have previously been evaluated within the frameworks of dimensional
regularisation \cite{deser:tmym, pisarski:tmym, mello:comment} and differential
regularisation \cite{chen:differential}, respectively. It was found that,
despite some controversy as to whether ambiguities might arise from the
continuation of the totally antisymmetric $\epsilon$-tensor to $d$ dimensions,
the results in both regularisation schemes agreed. Here, we have shown that the
self-energies of Chern-Simons massive three-dimensional Yang-Mills theory are
indeed finite and regularisation-independent for any gauge-invariant regulator.

With both parts of the self-energy at hand, we can now proceed to solve the gap
equation (\ref{gap}). Multiplying both sides by $m$ yields a value of
\begin{equation}
\label{mass}
m=\left(\frac{27}{16}\log 3-\frac{1}{4}\right)\frac{g_3^2N}{2\pi}\approx
1.604\frac{g_3^2N}{2\pi}
\end{equation}
for the magnetic mass. We note that the Ward identities of the theory
\cite{pisarski:tmym} ensure $\Im(A(-m^2))=\Im(B(-m^2))$ so that the
self-consistent mass will always be real.

\section{Discussion}

There have been several self-consistent studies of the magnetic mass using
parity-even quadratic mass terms:

Alexanian and Nair \cite{alexanian:magmass} have proposed a mass term that is
related to the generating functional for hard thermal loops, and have obtained
the mass
\begin{equation}
m_{AN}=\left(\frac{21}{16}\log 3 -\frac{1}{4}\right)\frac{g_3^2N}{2\pi}
\approx 1.192\frac{g_3^2N}{2\pi}
\end{equation}

Buchm\"uller and Philipsen \cite{buchmuller:magmass} have coupled a Higgs field
in the symmetric phase to the gluons and have found a gluon mass of 
\begin{equation}
m_{BP}=\left(\frac{63}{64}\log 3 - \frac{3}{16}\right)\frac{g_3^2N}{2\pi}
\approx 0.894\frac{g_3^2N}{2\pi}
\end{equation}
The results of Buchm\"uller and Philipsen have been extended to two loops by
Eberlein \cite{eberlein:twoloop}, who found
\begin{equation}
m_{Eb}\approx 1.052\frac{g_3^2N}{2\pi}
\end{equation}
which indicates a small, though non-negligible, contribution from higher loop
orders to the one-loop results.

Cornwall \cite{cornwall:magmass} has used a ``pinch'' technique for the
determination of the gluon mass and has obtained
\begin{equation}
m_{Co}=\left(\frac{15}{16}\log 3-\frac{1}{4}\right)\frac{g_3^2N}{2\pi}
\approx 0.780\frac{g_3^2N}{2\pi}
\end{equation}

A purely non-perturbative determination of the mass gap in (2+1)-dimensional
Yang-Mills theory has been carried out by Karabali, Kim and Nair
\cite{kkn:hamiltonian}. Using a
functional hamiltonian approach, they find that to lowest order in an enhanced
perturbative expansion, three-dimensional Yang-Mills theory is a theory of
massive interacting coloured particles with mass
\begin{equation}
m_{KKN}=\frac{g_3^2N}{2\pi}
\end{equation}

This agrees well with a recent lattice study by Philipsen
\cite{philipsen:nonperturbative}, who uses the mass splittings between $0^{++}$
and $1^{--}$ bound states of heavy scalars in the static limit to determine a
parton mass of
\begin{equation}
m_{P}= 0.360(19)g_3^2\sim 1.131(59)\frac{g_3^2N}{2\pi}
\end{equation}
for the three-dimensional SU(2) gluon.

The similarity between these non-perturbative results and the perturbative
calculations above lends credibility to the assumption that a magnetic screening
mass of order $\frac{g_3^2N}{2\pi}$ is generated for the chromomagnetostatic modes
in the QCD plasma. We also note the tantalising fact that the non-perturbative
result $m_{KKN}$ obeys the condition (\ref{quant}).

The result obtained with the Chern-Simons term is almost a factor of 2 larger
than those obtained using parity-even mass terms.

Compared to quadratic mass terms, the Chern-Simons term has the apparent
disadvantage of being odd under parity, thereby breaking the parity invariance
of the original theory. However, we do not necessarily consider this a
disadvantage, for the following reason:
The thermal ground state of the plasma is not the vacuum. Parity-odd terms made
up from chromomagnetic and -electric fields may have a non-vanishing expectation
value if something like a chromodynamic dynamo effect takes place in the plasma,
breaking the parity invariance of the action. The fact that in this case the
symmetry breaking necessary to have a non-vanishing 
Chern-Simons term in the action has to be caused be effects in the plasma, opens
the interesting possibility to have a domain structure within the plasma, where
different regions correspond to different signs of the Chern-Simons mass. If
this were the case, one would have to sum over both signs in calculating
long-distance contributions to thermodynamic quantities with a Chern-Simons
massive gluon. The existence of a domain structure would also imply the
existence of domain walls separating the two phases, which would have their own
(purely nonperturbative) mass scale associated with them, which would also have
to be taken into account in determining the long-range dynamics of the plasma.
The question of a possible domain structure in the plasma, and
possible observable consequences of parity breaking in the plasma certainly
deserve further research.

Another apparent problem lies in the
appearance of the $p^2$ term in the numerator of the propagator
(\ref{prop}), which as seen above leads to a branch point at $p^2=0$,
that seems to preclude exponential screening of
chromomagnetic fields. However, as noted by Deser, Jackiw and
Templeton \cite{deser:tmym}, and by Pisarski and Rao
\cite{pisarski:tmym}, no massless intermediate states may contribute
to the expectation values of gauge-invariant operators, so that the
physical long-range interactions would still be screened. In principle,
this could be demonstrated by computing the static quark potential,
which should not contain a massless contribution.

We conclude that the Chern-Simons term
cannot be excluded \emph{a priori} on the grounds that it is forbidden by
symmetry, and therefore must be included in the effective Lagrangian. From
there, our self-consistent calculation shows that it is possible to have a
non-zero value for the Chern-Simons mass. There are, however, many problems with
the Chern-Simons term as a mass term for the gluon, and further research is
certainly needed in order to arrive at any definite conclusions about its
possible role in chromomagnetic screening.

\section{Acknowledgments}

We gratefully acknowledge helpful comments by and discussions with I.T. Drummond
and A.K. Rajantie.

\end{document}